\documentclass[12pt]{amsart}
\usepackage{amsmath}
\usepackage{amsthm}
\usepackage{amsfonts}
\usepackage{graphicx}
\newenvironment{nouppercase}{
  
  \renewcommand{\uppercasenonmath}[1]{}}{}

\begin{document}

\title[On the r--matrix of M(embrane)--theory]
{On the r--matrix of M(embrane)--theory}
\author{Jens Hoppe}
\address{Braunschweig University, Germany}
\email{jens.r.hoppe@gmail.com}

\begin{abstract}
Supersymmetrizable theories, such as 
M(em)branes and associated matrix--models
related to Yang--Mills theory, possess r--matrices
\end{abstract}

\begin{nouppercase}
\maketitle
\end{nouppercase}
\thispagestyle{empty}
\noindent
While the Lax--pairs found in \cite{1,2}, in contrast to
standard integrable systems (see e.g. \cite{3}), naively
do not seem to provide any non--trivial conserved
quantity, it is also unlikely that they will not be useful.
As a start I would like to point out that Lax--pairs
arising from supersymmetrizability generically do have
an r--matrix associated with them (which in principle
is not even particularly difficult to explicitly calculate),
including the infinite--dimensional case of relativistic
higher dimensional extended objects (see e.g. \cite{4} for a review) such as Membrane theory,
whose discretized version, a SU(N)--invariant
matrix--model \cite{5}, is known to be subtle
in several ways (e.g. possessing classical solutions
extending to infinity, but quantum--mechanically
purely discrete spectrum \cite{6,7}, while when
supersymmetrized \cite{8} changing ``again'' to continuous\footnote{by many first interpreted negatively, then \cite{12,13} positively} \cite{9,10});
in this sense making the existence of a rather special
Lax--pair for them not too surprising.\\[0.15cm]
Let me first illustrate the idea by considering\footnote{cp.\cite{1}(missing $\frac{1}{2}$ in eq.(12), dWHN: 305/1988).} 
\begin{equation}\label{eq1} 
\begin{split}
\dot{L}_1 = [L_1, M], & \quad \dot{L}_2 = [L_2, M] \\
L_1 = \sum^N_{a=1}(\gamma_a p_a - \gamma_{a+N}\partial_a w), & \quad 
L_2 = \sum_a(\gamma_a \partial_a w + \gamma_{a+N} p_a)\\
M = -\frac{1}{2} & \sum^N_{a,b=1} \gamma_a \gamma_{b+N} \partial^2_{ab}w
\end{split}
\end{equation}
where $w = w(x_1, x_2, \ldots, x_N)$ and the hermitean Clifford matrices $\gamma_{i=1\ldots 2N,}$ satisfying
\begin{equation}\label{eq2} 
\gamma_i \gamma_j + \gamma_j \gamma_i = 2 \delta_{ij} \cdot \mathbf{1},
\end{equation}
canonically realized as $2^N \times 2^N$ dimensional tensorproducts of Pauli--matrices. While in that canonical representation $L_1$ and $L_2$ anticommute and square to a multiple of the unit matrix, $\beta,\, \beta' = 1, 2$, 
\begin{equation}\label{eq3} 
L_{\beta} L_{\beta'} + L_{\beta'} L_{\beta} = 2 \delta_{\beta \beta'}(2H := \vec{p}\,^2 + (\nabla w)^2) \mathbf{1}
\end{equation}
(\ref{eq1}), due to the polynomials of degree $\leq 2$ in the $\gamma_i$ closing under commutation, forming a (spinor--) representation of $so(2N+1)$, may also be considered in the {\it defining}, `vector' representation of $so(2N+1)$, in which 
$\tilde{L}(\lambda) := \frac{1}{2}(\tilde{L}_1 + \lambda \tilde{L}_2)$ and $\tilde{M}$, instead of having non--zero elements distributed over many of the $2^N \times 2^N$ entries, take the simple form
 \begin{equation}\label{eq4} 
\begin{split}
\tilde{L}(\lambda) = & \: i 
\begin{pmatrix}
0 & v^T \\ -v & 0_{2N \times 2N} 
\end{pmatrix}
=: \sqrt{2H} \sqrt{\lambda^2 + 1} K(\lambda)\\
v = & \: 
\begin{pmatrix}
\vec{p} + \lambda \vec{\nabla} w \\ \lambda \vec{p} - \vec{\nabla}w 
\end{pmatrix}
=: \sqrt{2H} \sqrt{\lambda^2 + 1} \;e(\lambda)\\
\tilde{M} = & 
\begin{pmatrix}
0 & 0 \\ 0 & -A
\end{pmatrix}, \quad
A = 
\begin{pmatrix}
0 & +w_{ab} \\ -w_{ab} & 0
\end{pmatrix}_{2N \times 2N},
\end{split}
\end{equation}
as when representing $\frac{1}{2} \gamma^{ij} = \frac{1}{4}(\gamma^i\gamma^j - \gamma^j\gamma^i)$ by 
$M_{ij} := E_{ij} - E_{ji}$ (generating $so(2N) \subset so(2N+1)$) $\frac{i\gamma_k}{2}$ will correspond to the generators $M_{0k} = E_{0k} - E_{k0}$ of $so(2N+1)$,
\begin{equation}\label{eq5} 
[M_{\mu \nu}, M_{\rho \lambda}] = \delta_{\nu \rho} M_{\mu \lambda} \pm 3 \text{more}
\end{equation}
$\mu,\nu,\rho,\lambda = 0,1,\ldots, 2N$.\\[0.15cm]
It is trivial to check that
\begin{equation}\label{eq6} 
\dot{\tilde{L}}(\lambda) = [\tilde{L}(\lambda), \tilde{M}] \Leftrightarrow 
\dot{e}(\lambda) = Ae, \quad e \in S^{2N-1}
\end{equation}
are equivalent to the equations of motion $\dot{x}_a = p_a$, $\dot{p}_a = -w_{ab}\partial_b w$, $(w_{ab} := \partial^2_{ab}w)$ being the Hessian of the `superpotential' $w$. However, in contrast with $L_1$ and $L_2$ in the spinor representation each having $N$ eigenvalues $+\sqrt{2H}$ and $N$ eigenvalues $-\sqrt{2H}$, the Lax--matrix $\tilde{L}(\lambda)$, as given in (\ref{eq4}), will have only {\it two} non--zero eigenvalues (which, diving by $\sqrt{2H}\sqrt{1+\lambda^2}$, i.e. considering the normalized matrix $K(\lambda)$, may be taken to be $\pm 1$), with corresponding eigenvectors $\hat{e}_{\pm}$, i.e.
\begin{equation}\label{eq7} 
\begin{split}
K(\lambda) & = U(\lambda)
\begin{pmatrix}
1 &  & & & \\
  &-1& & & \\
  &  &0& & \\
  &  & &\ddots & \\
  &  & & & 0
\end{pmatrix}
U^{\dagger}(\lambda),\\
U(\lambda) & = \frac{1}{\sqrt{2}}
\begin{pmatrix}
+i & -i & 0           & \ldots &0\\
e   & e   & \sqrt{2}n_1 & \ldots & \sqrt{2}n_{2N-1}
\end{pmatrix}
=(u_0 u_1 \ldots u_{2N-1})
\end{split}
\end{equation}
with $(e, n_1,\ldots, n_{2N-1})$ forming an orthonormal basis of $\mathbb{R}^{2N}$.\\[0.15cm] 
As the eigenvalues of $K(\lambda)$ are numerical constants $(= +1,-1,0 \ldots 0)$,hence Poisson--commuting with everything, it easily follows, with
\begin{equation}\label{eq8} 
\lbrace X_1, Y_2 \rbrace := \lbrace X \otimes \mathbf{1}, \mathbf{1} \otimes Y \rbrace := \lbrace X_{ij}, Y_{kl} \rbrace E_{ij}\otimes E_{kl}
\end{equation}
that $K_1 := K \otimes \mathbf{1}$ and $K_2 := \mathbf{1} \times K$ satisfy
\begin{equation}\label{eq9} 
\begin{split}
\lbrace K_1(\lambda), K_2(\lambda) \rbrace & = \big[ [U_{12}(\lambda), K_1 ], K_2 \big] = \big[ [U_{12}(\lambda), K_2 ], K_1 \big]\\
 & = [\mathring{r} _{12}, K_1] - [\mathring{r}_{21}, K_2]\\
U_{12} & := \lbrace U_1, U_2 \rbrace U^{-1}_1 U^{-1}_2 \\
\mathring{r} _{12} & = \frac{1}{2}[U_{12}, K_2], \: \mathring{r}_{21}  = \frac{1}{2}[U_{21}, K_1] = -\frac{1}{2}[U_{12},K_1];
\end{split}
\end{equation}
and $J(\lambda) := \sqrt{H}K(\lambda)$ will therefore satisfy
\begin{equation}\label{eq10} 
\begin{split}
\lbrace J_1, J_2 \rbrace & = H \lbrace K_1, K_2 \rbrace + \frac{1}{2} (\dot{K}_1 K_2 - \dot{K}_2 K_1)\\
 & = [H\mathring{r}_{12} - \frac{1}{2} \tilde{M}_1 K_2, K_1 ] - (1\leftrightarrow 2) 
\end{split}
\end{equation}
so that the $r$--matrix for the Lax--pair $(J(\lambda),\, \tilde{M} =
\big(
\begin{smallmatrix}
0 & 0 \\ 0 & A
\end{smallmatrix}
\big)
)$
with equations of motion $\dot{e} = Ae$, $e(\lambda) \in S^{2N-1}$ is 
\begin{equation}\label{eq11} 
r_{12}(\lambda) = \sqrt{H} \mathring{r}_{12}(\lambda)- \frac{1}{2}\tilde{M}_1 K_2(\lambda)\frac{1}{\sqrt{H}}. 
\end{equation}
Having gone through this simple example it is almost correct to say that one has understood {\it all} supersymmetrizable systems (from this perspective) whose supercharges are linear in the Clifford generators, resp. `fermions' (and at this stage it would be tempting to see a relation to other, old and new - see e.g. \cite{11}, and references therein - statements about supersymmetric systems); one `only' gets different (for field--theories: infinite dimensional) unit vectors $e$ and different (`more complicated') antihermitean operators $A$ (cp. (\ref{eq6})) satisfying 
$
\dot{e}(\lambda) = A e (\lambda) 
$.
\\[0.15cm] 
Consider now the membrane--matrix model (cp.(\ref{eq2})$_{J_{a=0}}$ of \cite{1}), i.e.
\begin{equation}\label{eq12} 
v = 
\begin{pmatrix}
\sum_{\beta} \lambda_{\beta} P^{\beta}_{\alpha a} \\
\sum_{\beta} \lambda_{\beta} Q^{\beta}_{\alpha a}
\end{pmatrix} =
\begin{pmatrix}
\sum_{\beta} \lambda_{\beta} \vec{p}\,^{\beta} \\
\sum_{\beta} \lambda_{\beta} \vec{q}\,^{\beta}
\end{pmatrix} =
\begin{pmatrix}
\vec{p}{(\lambda)} \\
\vec{q}{(\lambda)}
\end{pmatrix} =
\begin{pmatrix}
\vec{p} \\
\vec{q}
\end{pmatrix}
\end{equation}
\begin{equation*}
P^{\beta}_{\alpha a} = \sum^d_{t=1} p_{ta} \gamma^t _{\beta \alpha}, \quad
Q^{\beta}_{\alpha a} = \frac{1}{2}(\gamma^{st})_{\beta \alpha} f_{abc}x_{sb}x_{tc},
\end{equation*}
where the $f_{abc}$ are totally antisymmetric (real) structure constants of $su(N)$, $a,b,c = 1 \ldots N^2-1$, the $x_{sb}$ and 
$p_{tc}$ are canonically conjugate variables, the $\gamma^t$ are real symmetric $\sigma \times \sigma$ matrices satisfying 
$\gamma^s \gamma^t + \gamma^t \gamma^s = \delta^{st} \mathbf{1}$,
the time--evolution is given by 
\begin{equation}\label{eq13} 
H = \frac{1}{2}(P^{\beta}_{\alpha a}P^{\beta}_{\alpha a} + Q^{\beta}_{\alpha a}Q^{\beta}_{\alpha a})
= \frac{1}{2}(\vec{p}\,^{\beta} \vec{p}\,^{\beta} + \vec{q}\,^{\beta} \vec{q}\,^{\beta}),
\end{equation}
which is independent of $\beta$, sum over $(\alpha a) = (11)\ldots (\sigma, N^2-1$) and
\begin{equation}\label{eq14} 
J_a = f_{abc} x_{sb}p_{sc} \stackrel{!}{=} 0, 
\end{equation}
which also implies $v^T v = (2H)(\vec{\lambda}\,^2)$; the equations of motion can be written in the form (cp.(6) of \cite{1})
\begin{equation}\label{eq15} 
\dot{\vec{q}} = \Omega \vec{p},\quad \dot{\vec{p}} = \Omega \vec{q},
\end{equation} 
and the Lax--pair \cite{1}, when going to the defining vector representation of $so(2n+1)$, $n = \sigma(N^2 -1) \in \mathbb{N}$, becomes --as explained above--
 
\begin{equation}\label{eq16} 
\begin{split}
J(\lambda) &= i 
\begin{pmatrix}
0 & v^T \\ -v & 0
\end{pmatrix}
\frac{1}{\sqrt{2\vec{\lambda}\,^2}}\\
\tilde{M} & = 
\begin{pmatrix}
0 & 0 \\ 0 & -A
\end{pmatrix}, \:
A_{2n \times 2n}  =
\begin{pmatrix}
0 & \Omega \\ \Omega & 0
\end{pmatrix} = -A^T\\
\Omega_{\alpha a, \alpha' a'}  & = f_{aa'c} x_{tc}\gamma^t_{\alpha \alpha'} 
\end{split}
\end{equation}
with 
\begin{equation}\label{eq17} 
\begin{split}
J & = U
\begin{pmatrix}
\sqrt{H} &  & & & \\
  &-\sqrt{H}& & & \\
  &  &0& & \\
  &  & &\ddots & \\
  &  & & & 0
\end{pmatrix}
U^{\dagger}\\
U & = 
\begin{pmatrix}
\frac{i}{\sqrt{2}} & -\frac{i}{\sqrt{2}} & 0           & \ldots &0\\
\frac{e}{\sqrt{2}}  & \frac{e}{\sqrt{2}}  & n_1 & \ldots & n_{\ast}
\end{pmatrix}
\end{split}
\end{equation}
and the $2n$ unit vectors $(e,n_1 \ldots n_{\ast})$ being orthonormal, and 
\begin{equation}\label{eq18} 
\dot{J}(\lambda) = [J(\lambda), \tilde{M}] \Leftrightarrow \dot{e} = Ae
\end{equation}
being equivalent to the matrix--model equations of motion.\\[0.15cm]
As shown above, the $su(N)$--invariant membrane matrix model of \cite{5} therefore possesses an r--matrix, 
\begin{equation}\label{eq19} 
\begin{split}
\lbrace J_1(\lambda), J_2(\lambda) \rbrace & = [r_{12}(\lambda),  J_1] - (1 \leftrightarrow 2)\\
r_{12} & = \big( \frac{1}{2} [U_{12}, J_2] - \frac{1}{2}\frac{\tilde{M}_1 J_2}{H} \big);
\end{split}
\end{equation}
note that the normalisation of $J$ is chosen such that $\frac{1}{2}\text{tr} J^2 = H$
(as a consistency check, one can calculate $-\text{Tr}_2(r_{12} J_2) = \frac{1}{2}\tilde{M}\frac{\text{tr}J^2_2}{H}
+ \frac{1}{4}\text{tr}_2[U_{12}, J^2_2]$
which indeed gives $\tilde{M}$; that $\mathring{r}_{12}$ does not give any contribution means that it is in some sense `trivial', i.e. not influencing the time--evolution; dimensionally 
$[x] \sim E^{\frac{1}{4}}$, $[p] \sim E^{\frac{1}{2}}$ so
$[\frac{\partial}{\partial x} \frac{\partial}{\partial p}] = E^{-\frac{3}{4}} = [\frac{\tilde{M}}{H}] = \frac{E^{\frac{1}{4}}}{E})$.\\[0.15cm] 
What about the infinite--dimensional case of membrane--theory?
\begin{equation}\label{eq20} 
\begin{split}
v^{\beta}_{\alpha a} & = \int Y_a(\varphi)\big(\frac{p_i}{\rho}\gamma^i_{\beta \alpha} + \frac{1}{2} \lbrace x_i, x_j \rbrace \gamma^{ij}_{\beta \alpha} \big) \rho\,d^2\varphi \quad a\in \mathbb{N}_0 \\
\lbrace x_i, x_j\rbrace (\varphi) & := \frac{\varepsilon^{rs}}{\rho} \partial_r x_i \partial_s x_j\\
\int Y_a Y_b \rho \, d^2 \varphi & = \delta_{ab},\quad 
\sum^{\infty}_{a=0} Y_a(\varphi) Y_a(\tilde{\varphi}) = \frac{\delta^2(\varphi, \tilde{\varphi})}{\rho},
\end{split}
\end{equation}
\begin{equation}\label{eq21} 
\begin{split}
\dot{J}(\lambda) & = [J(\lambda), \tilde{M}], \\
\Omega_{\alpha a, \alpha' a'} & = g_{aa'c} x_{ic}\gamma^i_{\alpha \alpha'},\:
g_{abc} = \int Y_a\lbrace Y_b, Y_c \rbrace \rho\,d^2\varphi\\
J &= i 
\begin{pmatrix}
0 & v^{\dagger} \\ -v & 0
\end{pmatrix}\frac{1}{\sqrt{2\lambda^{\dagger}\lambda}},\\
\tilde{M} & = 
\begin{pmatrix}
0 & 0 \\ 0 & -A
\end{pmatrix}, \:
A  =
\begin{pmatrix}
0 & \Omega \\ \Omega & 0
\end{pmatrix}.
\end{split}
\end{equation}
While in principle having to worry about potentially diverging infinite sums, and Lie--algebraically one would have to identify a well--defined algebra, I think that (\ref{eq21}), and the infinite--dimensional analogue of (\ref{eq19}), should be fine, for various reasons: 
$v^{\dagger} v = 2\lambda^{\dagger}\lambda H$ implies that for fixed (trivially conserved) energy all components of $v$ are finite, and 
$\frac{v}{\sqrt{2\lambda^{\dagger}\lambda H}} =: e$ will be a unit vector;
the norm of each row of $\Omega$ (or column, given by $(\alpha a)$) is $\int \lbrace Y_a, x_i \rbrace^2 \rho\, d^2\varphi$, which is clearly finite, as one integrates over a compact manifold, and if $\lbrace Y_a, x_i \rbrace$ was infinite for some $a$, the potential term of $H$ could not be finite; the arising scalar products of infinite--dimensional vectors (corresponding to $\int \rho\,d^2 \varphi$ of the product of corresponding square--integrable functions) therefore involve only vectors of finite norm. One may also write (\ref{eq21}), resp. the equations of motion, in the following compact suggestive forms:
\begin{equation*} 
\begin{split}
\dot{q}^{\beta}_{\alpha} & = \gamma^i_{\alpha \alpha'} \lbrace p^{\beta}_{\alpha'}, x_i\rbrace, \: 
\dot{p}^{\beta}_{\alpha}  = \gamma^i_{\alpha \alpha'} \lbrace q^{\beta}_{\alpha'}, x_i \rbrace, \: \text{or as}\\
\dot{V} & = \lbrace V, X \rbrace := \lbrace V_{\beta \alpha}, X_{\delta \varepsilon} \rbrace E_{\beta \alpha} E_{\delta \varepsilon} = -\lbrace X, V^T \rbrace^T
\end{split}
\end{equation*}
resp.
\begin{equation}\label{eq22} 
\begin{split}
\dot{Q}_{\beta \alpha} & = \lbrace P_{\beta \alpha'}, X_{\alpha'\alpha} \rbrace = - \lbrace X_{\alpha \alpha'}, P_{\alpha' \beta} \rbrace = -\dot{Q}_{\alpha \beta}\\
\dot{P}_{\beta \alpha} & = \lbrace Q_{\beta \alpha'}, X_{\alpha'\alpha} \rbrace = + \lbrace X_{\alpha \alpha'}, Q_{\alpha' \beta} \rbrace = +\dot{P}_{\alpha \beta}
\end{split}
\end{equation}
(using that, as finite matrices, $P$ is symmetric, $Q = \frac{1}{2}\lbrace X, X \rbrace$ antisymmetric and $X_{\alpha \alpha'} := \gamma^i_{\alpha \alpha'}x_i = X_{\alpha' \alpha}$, $\dot{X} = P$, $\ddot{X} = \frac{1}{2}\lbrace X, \lbrace X, X \rbrace \rbrace$).\\[0.15cm]
Finally, note \cite{15} and that $L = P+iQ$, respectively $\dot{L} = i\lbrace L^{\ast}, X \rbrace$ could be considered a generalization to arbitrary $d$ of (39) in \cite{14}, turning into a real Lax--pair for the Wick--rotated/Euclidean equation of motion (though care is needed for the definition of a Lie--algebra involving matrix--valued functions on the membrane).

\vspace{0.5cm}
\noindent
\textbf{Acknowledgement.} 
I would like to thank J. Fr\"ohlich and I. Kostov for discussion,
T. Ratiu for having raised the question on which Lie--algebra
the Lax--pair (12) of \cite{1} is formulated, and O. Lechtenfeld
for sending me a copy of his Ph.D. thesis.

\end{document}